\documentclass[12pt,oneside,peerreview,draftcls,onecolumn]{IEEEtran}
\usepackage{amsmath,amsfonts,amssymb,amsbsy, amsthm,ae,aecompl}
\usepackage{algorithm,algpseudocode}
\usepackage[utf8]{inputenc}
\usepackage[english]{babel}
\usepackage{bm}
\usepackage{color}
\usepackage{cite}
\usepackage{epsfig}
\usepackage{enumerate}
\usepackage{float}
\usepackage{fancyhdr}
\usepackage[T1]{fontenc}
\usepackage[acronym,toc,shortcuts]{glossaries}
\usepackage{graphicx, caption, subcaption}
\usepackage{hyperref}
\usepackage{bbm}
\usepackage{lastpage}
\usepackage{listings}
\usepackage{lipsum}
\usepackage{dsfont}
\usepackage{overpic}
\usepackage{transparent}
\usepackage{tikz,pgfplots}
\usepackage{times}
\usepackage{verbatim}
\usepackage[all]{xy}

\hypersetup{
    colorlinks,%
    citecolor=black,%
    filecolor=black,%
    linkcolor=black,%
    urlcolor=black
}

\usepackage{etoolbox}
\AtBeginEnvironment{align}{\setcounter{subeqn}{0}}
\newcounter{subeqn} %

\usetikzlibrary{calc}
\usetikzlibrary{arrows,decorations.markings}
\pgfplotsset{
  grid style = {
    dash pattern = on 0.025mm off 0.95mm on 0.025mm off 0mm, 
    line cap = round,
    black,
    line width = 0.5pt
  },
  tick label style={font=\small},
  label style={font=\small},
  legend style={font=\footnotesize},
}

\pgfplotscreateplotcyclelist{laneas-vr1}{
cyan!60!black,			solid, 	every mark/.append style={fill=cyan!60!black},mark=x\\%
cyan!60!black,			solid, 	every mark/.append style={fill=cyan!60!black},mark=x\\%
yellow!60!black,		solid, 	every mark/.append style={fill=cyan!60!black},mark=+\\%
yellow!60!black,		solid, 	every mark/.append style={fill=cyan!60!black},mark=+\\%
red!60!black,			solid, 	every mark/.append style={fill=cyan!60!black},mark=o\\%
red!60!black,			solid, 	every mark/.append style={fill=cyan!60!black},mark=o\\%
}
\graphicspath{{figures/}}

\bgroup
\makeglossaries
\newacronym{AR}{AR}{augmented reality}
\newacronym{AI}{AI}{artificial intelligence}
\newacronym{BS}{BS}{base station}
\newacronym{CBR}{CBR}{channel  busy  ratio}
\newacronym{CPU}{CPU}{central processing unit}
\newacronym{D2D}{D2D}{device-to-device}
\newacronym{IDR}{IDR}{information  dissemination  rate}
\newacronym{IVR}{IVR}{interconnected virtual reality}
\newacronym{IoT}{IoT}{Internet of Things}
\newacronym{MAC}{MAC}{media access control}
\newacronym{MEC}{MEC}{mobile edge computing}
\newacronym{NFV}{NFV}{network function virtualization}
\newacronym{RB}{RB}{resource block}
\newacronym{RAN}{RAN}{radio access network}
\newacronym{PER}{PER}{packet  error  rate}
\newacronym{SDN}{SDN}{software defined networking}
\newacronym{SRP}{SRP}{Ski-Rental problem}
\newacronym{UHD}{UHD}{ultra high definition}
\newacronym{UE}{UE}{user equipment}
\newacronym{VR}{VR}{virtual reality}

\begin{document}
\title{Towards Interconnected Virtual Reality: Opportunities, Challenges and Enablers}
\author{
		\IEEEauthorblockN{Ejder Baştuğ$^{\diamond, \otimes}$, Mehdi Bennis$^{\dagger}$, Muriel Médard$^{\diamond}$, and Mérouane Debbah$^{\otimes, \circ}$\\}
		\IEEEauthorblockA{
				\small
				$^{\diamond}$Research Laboratory of Electronics, Massachusetts Institute of Technology, \\ 77 Massachusetts Avenue, Cambridge, MA 02139, USA \\				
				$^{\otimes}$Large Networks and  Systems Group (LANEAS), CentraleSupélec, \\ Université Paris-Saclay, 3 rue Joliot-Curie,  91192 Gif-sur-Yvette, France \\	
				$^{\dagger}$Centre for Wireless Communications, University of Oulu, Finland \\	
				$^{\circ}$Mathematical and Algorithmic Sciences Lab, Huawei France R\&D, Paris, France \\
				\{ejder, medard\}@mit.edu, bennis@ee.oulu.fi, merouane.debbah@centralesupelec.fr
				\vspace{-0.65cm}
		}
		\thanks{This research has been supported by the ERC Starting Grant 305123 MORE (Advanced Mathematical Tools for Complex Network Engineering), the  U.S.  National  Science  Foundation  under  Grant CCF-1409228, and the Academy of Finland CARMA project. NOKIA Bell-Labs is also acknowledged for its donation to the "FOGGY" project.}
}
\IEEEoverridecommandlockouts
\maketitle

\begin{abstract}
Just recently,  the concept of augmented and virtual reality (AR/VR) over wireless has taken the entire 5G ecosystem by storm spurring an unprecedented interest from both academia, industry and  others. Yet, the success of an immersive VR  experience hinges on solving a plethora of grand challenges cutting across multiple disciplines. This article underscores the importance of VR technology as a disruptive use case of 5G (and beyond) harnessing the latest development of storage/memory, fog/edge computing, computer vision, artificial intelligence and others. In particular, the main requirements of wireless interconnected VR are  described followed by a selection of key enablers, then, research avenues and their underlying grand challenges are presented.  Furthermore, we examine three VR case studies and provide numerical results under various storage, computing and network configurations. Finally, this article exposes the limitations of current networks and makes the case for more theory, and innovations to spearhead VR for the masses.
\end{abstract}
\vspace{-0.25cm}
\section{Introduction}
We are on the cusp of a true revolution which will transcend everything we (humans) have witnessed so far. Leveraging recent advances in storage/memory, communication/connectivity, computing, big data analytics, \ac{AI}, machine vision and other adjunct areas will enable the fruition of immersive technologies such as augmented and virtual reality (AR/VR). These technologies will enable the transportation of ultra-high resolution light and sound in real-time to another world through  the relay of its various sights, sounds and  emotions. The use of \ac{VR} will go beyond early adopters such as gaming to enhancing cyber-physical and social experiences such as conversing with family, acquaintances, business meeting, or disabled persons. Imagine if one could put on a \ac{VR} headset and walk around a street where everyone is talking Finnish and interact with people in Finnish in a fully immersive experience.  Add to this the growing number of drones, robots and other self-driving vehicles taking cameras to places humans could never imagine reaching; we shall see a rapid increase of new content from fascinating points of view around the globe. Ultimately \ac{VR} will provide the most personal experience with the closest screen, providing the most connected, most immersive experience witnessed thus far.

\Ac{AR} and \ac{VR} represent two ends of the spectrum. One the one hand \ac{AR}  bases reality as the main focus and the virtual information is presented over the reality, whereas \ac{VR} bases virtual data as the main focus, having the user immerse into the middle of the synthetic reality virtual environment. One can also imagine a mixed reality where \ac{AR} meets VR, by merging the physical and virtual information seamlessly. Current online social networking sites (Facebook, Twitter, and the likes) are just precursors of what we will come to truly witness when social networking will encompass immersive virtual-reality technology. At its most basic, social virtual reality allows two geographically separated people (in the form of avatars) to communicate as if they were face-to-face. They can make eye contact and can manipulate virtual objects that they both can see. Current \ac{VR} technology is in its inception since headsets are not yet able to track exactly where eyes are pointed at, by instead looking at the person one is talking at. Moreover, current state-of-the-art \ac{VR} technology is unable to read detailed facial expressions and senses. Finally, and perhaps the biggest caveat is that most powerful \ac{VR} prototypes are wired with cables because the amount of transmitted high-resolution video at high frame rates simply cannot be done using today's wireless technology ($4$G/LTE), let alone the fact that a perfect user interface (\ac{VR} equivalent of the mouse) is still in the making. 
\begin{figure*}[ht!]
	\centering
	\includegraphics[width=0.95\linewidth]{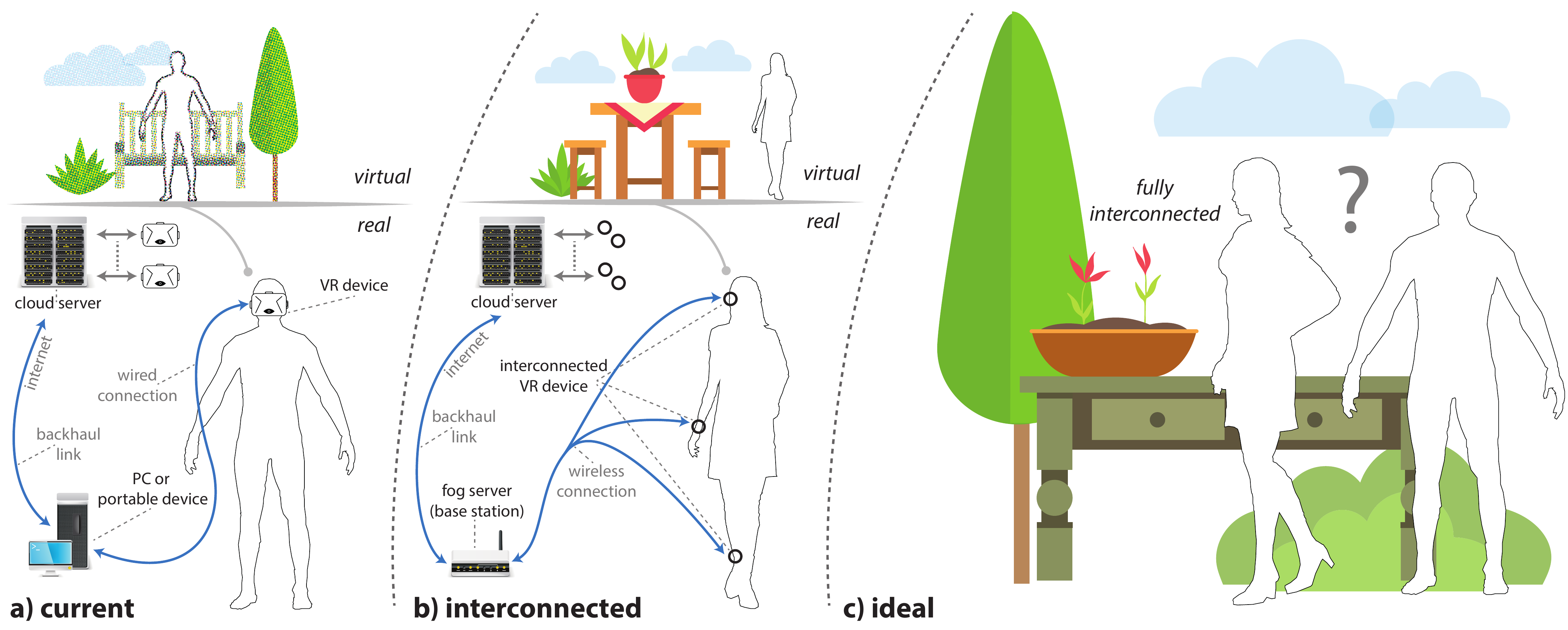}
	\caption{An illustration of virtual reality scenarios: a) current virtual reality systems, b) interconnected, and c) ideal (fully interconnected) systems.}
	\label{fig:generalscenario}
\end{figure*} 

These shortcomings have spurred efforts to make social \ac{VR} happen in the near future. A number of startup companies such as Linden Lab (a screen-based simulation) is getting ready to roll out a new platform called SANSAR \cite{ProjectSansar} which is a host for user-created virtual experiences and tools for \ac{VR} headsets, standard computer monitors, and mobile devices. Similarly, the SANSAR world will function much like Second Life, with people leasing space for their virtual creations, rendered in 3D and at a high frame rate. Likewise, BELOOLA \cite{Beloola} is building a virtual world designed for social networking. These recent trends are a clear indication that the era of responsive media is upon us, where media prosumers will adapt content dynamically to match consumers' attention, engagement and situation. While some of the \ac{VR} technologies are already emerging (\ac{VR} goggles, emotion-sensing algorithms, and multi-camera systems), current $4$G (or even pre-5G) wireless systems {\bf cannot cope} with the massive amount of bandwidth and latency requirements of \ac{VR}. 

The goal of this article is to discuss current and future trends of \ac{VR} systems, aiming at reaching a fully interconnected \ac{VR} world. It is envisaged that \ac{VR} systems will undergo  three different evolution stages as depicted in Fig. \ref{fig:generalscenario}, starting with current \ac{VR} systems, evolving towards \ac{IVR}, and finally ending up with the ideal \ac{VR} system.  The rest of this paper is dedicated to a discussion of this evolution, laying down some of the key enablers and requirements for the ultimate \ac{VR} technology. In this regard, we discuss current \ac{VR} systems and limits of human perception in Section \ref{sec:twoard}, prior to shifting towards interconnected \ac{VR} and related technological requirements.  Key research avenues and scientific challenges are detailed in Section \ref{sec:research}. Several case studies (with numerical results) are given in Section \ref{sec:numresults}. Finally, Section \ref{sec:idealvr} debates whether an ideal fully-interconnected \ac{VR} system can be achieved and what might be needed in this regard.
\section{Toward Interconnected VR}\label{sec:twoard}
\begin{figure}[ht!]
 	\centering
	\includegraphics[width=1.0\linewidth]{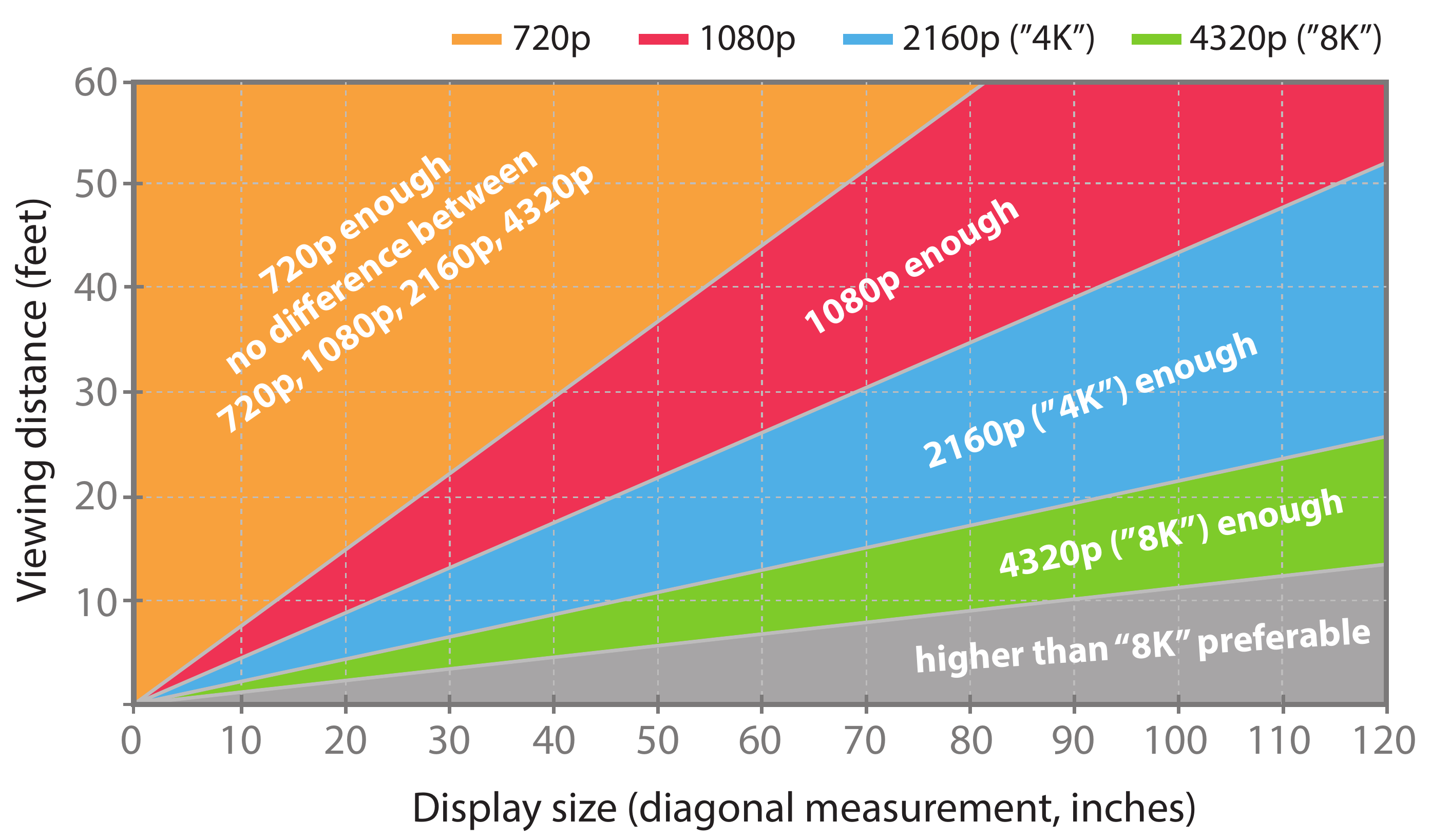}
	\caption{Display size versus viewing distance (see \cite{Size2Distance} for an interactive example).}
	\label{fig:sizeview} 
\end{figure}

The overarching goal of virtual reality is to generate a digital real-time experience which mimics the full resolution of human perception. This entails recreating every photon our eyes see, every small vibration our ears hear and other cognitive aspects (e.g., touch, smell, etc.). Quite stunningly, humans process nearly $5.2$ gigabits per second of sound and light. The fovea of our eyes can detect fine-grained dots allowing them to differentiate approximately $200$ distinct dots per degree (within our foveal field of view) \cite{Konig1897Abhangigkeit, Pirenne1967Vision}. Converting that to pixels on a screen depends on the size of the pixel and the distance between our eyes and the screen, while using $200$ pixels per degree as a reasonable estimate (see Fig. \ref{fig:sizeview} for an estimate). Without moving the head, our eyes can mechanically shift across a field of view of at least $150$ degrees horizontally (i.e., $30.000$ pixels) and $120$ degrees vertically (i.e., $24.000$ pixels). This means the ultimate \ac{VR} display would need a region of $720$ million pixels for full coverage. Factoring in head and body rotation for $360$ horizontal and $180$ vertical degrees amounts to a total of more than $2.5$ billion (Giga) pixels. Those are just for a static image. 

For motion video, multiple static images are flashed in sequence, typically at a rate of $30$ images per second (for film and television). But the human eye does not operate like a camera. Our eyes actually receive light constantly, not discretely, and while $30$ frames per second is adequate for moderate-speed motion in movies and TV shows, the human eye can perceive much faster motion ($150$ frames per second). For sports, games, science and other high-speed immersive experiences, video rates of $60$ or even $120$ frames per second are needed to avoid motion blur and disorientation. Assuming no head or body rotation, the eye can receive $720$ million pixels for each of $2$ eyes, at $36$ bits per pixel for full color and at $60$ frames per second, amounting to  a total of $3.1$ trillion (tera) bits! Today's compression standards can reduce that by a factor of $300$ and even if future compression could reach a factor of $600$ (the goal of future video standards), that still means $5.2$ gigabits per second of network throughput (if not more) is needed. While $8$K cameras are being commercialized no cameras or displays to date today can deliver $30$K resolution. 

As a result,  media prosumers are no longer using just a single camera to create experiences. At today's $4$K resolution, $30$ frames per second and $24$ bits per pixel, and using a $300:1$ compression ratio, yields $300$ megabits per second of imagery. That is more than $10$x the typical requirement for a high-quality $4$K movie experience. While panorama camera rigs face outward, there is another kind of system where the cameras face inward to capture live events. This year's Super Bowl, for example, was covered by $70$ cameras $36$ of which were devoted to a new kind of capture system which allows freezing an action while the audience pans around the center of the action. Previously, these kinds of effects were only possible in video games because they require heavy computation to stitch the multiple views together. A heavy duty post-processing means such effects are unavailable during live action.  

\textbf{As  a result, $5$G network architectures are being designed to move the post-processing at the network edge so that processors at the edge and the client display devices (\ac{VR} goggles, smart TVs, tablets and phones) carry out advanced image processing to stitch camera feeds into dramatic effects.}

To elaborate the context of current networks, even with a dozen or more cameras capturing a scene, audiences today only see one view at a time. Hence, the bandwidth requirements would not suffice to provide an aggregate of all camera feeds. To remedy to this, dynamic caching and multicasting may help alleviate the load, by delivering content to thousands from a single feed. In a similar vein with the path towards \ac{UE} centricity \ac{VR} will instead let audiences dynamically select their individual point of view. That means that the feed from \emph{all} of the cameras needs to be available instantly and at the same time, meaning that conventional multicast will not be possible when each audience member selects an individualized viewpoint (unicast). This will cause outage and users' dissatisfaction.
%
\subsection{Technological Requirements}\label{sec:treq} 
In order to tackle these grand challenges, the $5$G network architecture (\ac{RAN}, Edge and Core) will need to be much smarter than ever before by adaptively and dynamically making use of concepts such as \ac{SDN}, \ac{NFV} and network slicing, to mention a few facilitating a more flexible allocating resources (\glspl{RB}, access point, storage, memory, computing, etc.) to meet these demands. In parallel to that video/audio compression technologies are being developed to achieve much higher compression ratios for new multi-camera systems. Whereas conventional video compression exploits the \emph{similarity of the images} between one frame and the next (\emph{temporal redundancy}), \ac{VR} compression adds to that and leverages similarity among images from different cameras (like the sky, trees, large buildings and others, called \emph{spatial redundancy}) and use intelligent slicing and tiling techniques, using less bandwidth to deliver full $360$ degree video experiences. All of these advances may still not be enough to reach the theoretical limits of a fully immersive experience. Ultimately, a fundamentally new network architecture is desperately needed that can dynamically multicast and cache multiple video feeds close to consumers and perform advanced video processing within the network to construct individualized views. 

Immersive technology will require massive improvements in terms of {\bf bandwidth}, {\bf latency} and {\bf reliability}. Current remote-reality  prototype (MirrorSys \cite{Mirrorsys}) requires $100$-to-$200$Mbps for a one-way immersive  experience. While MirrorSys uses a single $8$K, estimates about photo-realistic \ac{VR} will require two $16$K $\times$ $16$K screens (one to each eye). Latency is the other big issue in addition to reliability. With an augmented reality headset, for example, real-life visual and auditory information has to be taken in through the camera and sent to the fog/cloud for processing, with digital information sent back to be precisely overlaid onto the real-world environment, and all this has to happen in less time than it takes for humans to start noticing lag (no more than $13$ms \cite{Potter2014Detecting}).  Factoring in the much needed high reliability criteria on top of these bandwidth and delay requirements clearly indicates the need for interactions between several research disciplines. These research avenues are discussed in the following.
%
\section{Key Research Avenues and Scientific Challenges}\label{sec:research}
\begin{figure}[ht!] 
	\centering
	\includegraphics[width=1.0\linewidth]{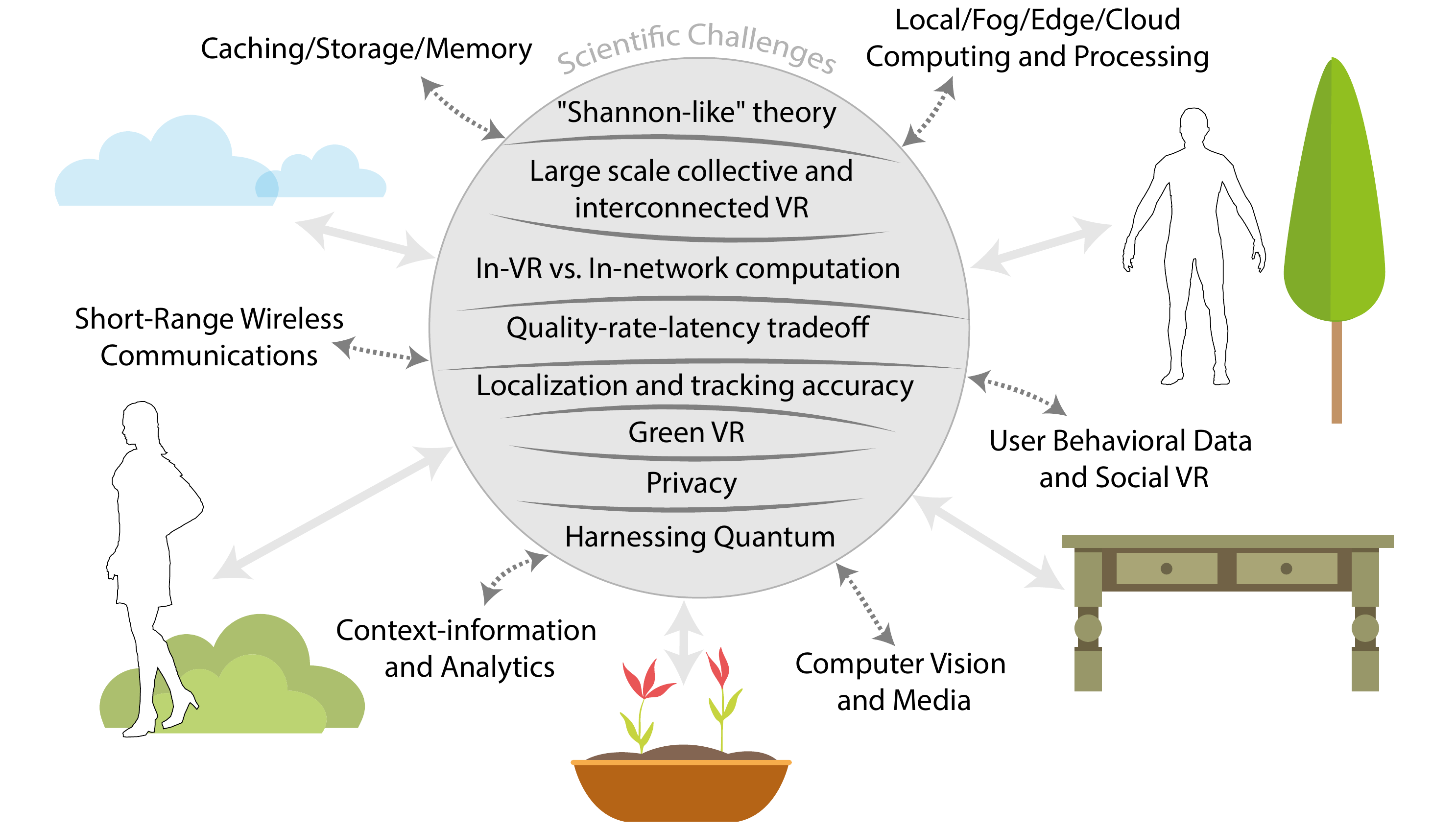}
	\caption{Research avenues and scientific challenges for interconnected \ac{VR}.}
	\label{fig:challenges}
\end{figure}

The success of interconnected \ac{VR} hinges on solving a number of research and scientific challenges across network and devices with heterogeneous capability of  storage, computing, vision, communication and context-awareness. These key  research directions  and scientific challenges are summarized in Fig. \ref{fig:challenges}, and discussed as follows.
%
\subsection{Caching/Storage/Memory}
The concept of content caching has been recently investigated in great details \cite{Bastug2014LivingOnTheEdge}, where the idea is to cache strategic contents at the network edge (at \ac{BS}, devices or other intermediate locations). One distinguishes between reactive and proactive caching. While the former serves end users when they request contents, the latter is  proactive and anticipates users' requests. Proactive caching  depends on the availability of fine-grained spatio-temporal traffic predictions. Other side information such as user's location, mobility patterns and social ties can be further exploited especially when context information is sparse. Storage will play a crucial role in \ac{VR} where for instance upon the arrival of a task query, the network/server needs to swiftly decide whether to store the object if the same request will come in the near future or instead recompute the query from scratch if the arrival rate of the queries will be sparse in the future. Content/media placement and delivery will also be important in terms of storing different qualities of the same content at various network locations \cite{Bethanabhotla2015Adaptive, Paschos2016Wireless}.
%
\subsection{Local/Fog/Edge/Cloud Computing and Processing}
\label{sec:research:computing}
Migrating computational intensive tasks from \ac{VR} devices to more resourceful cloud/fog servers is necessary to increase the computational capacity of low-cost devices while saving battery energy.  For this purpose, \Ac{MEC} will enable devices to access cloud/fog resources (infrastructures, platforms, and software)  in an on-demand fashion. While current state of the art solutions allocate radio and computing resources in a centralized manner (at the cloud), for \ac{VR} both radio access and computational resources must be brought closer to \ac{VR}-users by harnessing the availability of dense small cell base stations with  proximity access to computing/storage/memory resources.  Furthermore, the network infrastructure must enable a fully distributed cloud immersive experience where a lot of the computation happens on very powerful servers that are in the cloud/edge while sharing the sensor data that are being delivered by end-user devices at the client side. In the most extreme cases, one can consider the computation at a very local level, say with fully/partially embedded devices in the human body, having computing capabilities. This phenomenon is commonly referred to   as ``skin computing''.
%
\subsection{Short-Range Wireless Communications} 
Leveraging short- range communication such as \ac{D2D} and edge proximity services among collocated \ac{VR}-users can help alleviate network congestion. The idea is to extract, stitch and share relevant contextual information among \ac{VR} users in terms of views and camera feeds. In the context of self-driving vehicles equipped with \ac{UHD} cameras  capturing their local neighborhood, the task for the vehicle/robot is to not only recognize objects/faces in real time but also decide which objects should be included in the map and share it with nearby vehicles for richer and more context-aware maps.
%
\subsection{Computer Vision and Media}
The advent of \ac{UHD} cameras ($8$K, new cameras with $360$-degree panoramic video) has enriched new video and media experiences. At the same time today's media content sits at two extreme ends of a spectrum. On the one hand one distinguishes "lean-back experiences" such as movies and television where consumers are passive and are led through a story by content authors/producers. On the other hand are "lean-forward" experiences in the form of games in which the user is highly engaged and drives the action through an environment created by content authors/producers. The next generation of "interactive media" where the narrative can be driven by authors/producers will be tailored dynamically to the situation and preferences of audience and end-users. 
%
\subsection{Context-information and Analytics}
Use of context-information has already been advocated as a means of optimizing complex networks. Typically, context-information refers to in-device and in-network side information (user location, velocity, battery level and other MAC/high layer aspects). In the context of \ac{VR}, the recent acquisition of Apple of Emotient, a company  using advanced computer vision to recognize the emotions of people serves as a clear indication that context-information will play an ever instrumental role in spearheading the success of \ac{VR}. In order to maximize the user's connected and immersive experience the emotional, user switchiness and other behavioral aspects must be factored in. This entails \emph{predicting users disengagement} and preventing it by dynamically shifting the content to better match individual's preferences, emotion state and situation. Since a large amount of users data in the network can be considered for the big-data processing,  tools from machine learning can be exploited to infer on the context-information of users and act accordingly. Of particular importance is the fact that  deep learning models have been recently on rise in machine learning applications, due to their human-like behavior in training and good performance in feature extraction.
%
\subsection{User Behavioral Data and Social VR}
A by-product  of the proliferation of multiple screens, is the notion of \emph{switchiness} is more prevalent in which users' attention goes from one screen to another. Novel solutions based on user behavioral data and social interactions must be thought off to tackle user's switchiness. For this purpose, the switchiness and  \emph{screen chaos} problems have basically the same answer. An immersive experience is an integrated experience which needs a data-driven framework that takes all of the useful information a person sees and bring it to a single place. Today that integration does not happen because there is no common platform. \ac{VR} mandates that all of these experiences take place in one place. If one is watching a movie, or playing a game, and get a phone call, the game (or movie) is automatically paused and the person need not have to think about pausing the movie and answering the phone. Considering that a common data-driven platform is taking in one place,  big data and machine learning tools will play a crucial role in  bringing the immersive experience to the users.
\subsection{Scientific Challenges}
\label{sec:scientific}
The goal of this subsection is to lay down the foundations of \ac{VR}, by highlighting the key different research agenda and potential solution concepts for its success.

{\bf Need for a  ``Shannon-like'' theory.} For a given \ac{VR} device of $S$ bits of storage, $E$ joules of energy and $C$ hertz of processing power, how to maximize the user's immersive experience or alternatively minimize \ac{VR}-users' switchiness?. The answer  depends on many parameters such as the \ac{VR} device-server air link, whether the \ac{VR} device is a  human or a robot, network congestion, in-\ac{VR} processing,  \ac{VR} cost (how much intelligence can be put at the \ac{VR} headset), distinction between massive amount of \ac{VR} devices transmitting few bits versus few of them sending ultra-high definition to achieve a specific task. In this regard, haptic code design for \ac{VR} systems, code construction to minimize delay in feedback scenarios \cite{Lucani2012Coding}, source compression under imperfect knowledge of input distribution, and granularity of learning the input distribution in source compression, become relevant. Moreover, Nyquist sampling with no prior knowledge, compressed sensing with partial structural knowledge, and source coding with complete knowledge are some of key scientific venues which can address many challenges in \ac{VR} networks.

{\bf Large scale collective and interconnected \ac{VR}.} The analysis of very large \ac{VR} networks and systems, most of them moving, is also of high interest. With so many different views and information, lots of redundancy and collective intelligence  is open to exploitation for the interconnected VR. 

{\bf  In-\ac{VR} vs. In-network computation}. This refers to where and to which level should the decoupling between in-\ac{VR} headset  and in-network computing happen. This depends on the bandwidth-latency-cost-reliability tradeoffs, where computing for low-end and cheap headsets needs to happen at the network-side, whereas for more sophisticated \ac{VR} headsets computing can be carried locally. 

{\bf  Quality-rate-latency tradeoff.} Given an underlying network topology, storage and communication constraints, what is the quality level per content that should be delivered to maximize the quality of an immersive \ac{VR} experience?  This builds on the works of Bethanabhotla et. al. \cite{Bethanabhotla2015Adaptive}   by  taking into account the video size and quality as a function of the  viewing distance.  Moreover, for a given latency, rate constraints  what is the optimal payload size for a given content to maximize information dissemination rate (in case of self-driving vehicles). Moreover, machine learning is key for object recognition and stitching different video feeds.  For self-driving vehicles, given an arbitrary number of vehicles, network congestion and wireless link among vehicles, \ac{CPU}, storage constraint and  vehicles aiming at  exchanging their local maps.  Fundamentally speaking, for a fixed packet size of $L$ bits, what objects need to be recognized/quantized and included in the map? for e.g., the map should store popular objects that have been requested a lot in the past.

{\bf Localization and tracking accuracy.}  For a fully immersive \ac{VR} experience,  very accurate localization techniques are needed, including the positions of objects, tracking of human eyes (i.e., gaze tracking) and so on.

{\bf Green \ac{VR}.} For a given target \ac{VR}-user's immersive experience, the goal is to minimize the power consumption in terms of storage, computing and communication. With the green interconnected \ac{VR}, the notion of ``charging'' the equipment should disappear/minimized, since this operation does not exist in the virtual world. Therefore, smart mechanisms for seamless charging of \ac{VR} devices (i.e., wireless power transfer/charging and energy harvesting) are  promising.

{\bf Privacy}. With users contributing to the world with different contents and having multiple views from billions of objects and users, the issue of privacy naturally takes central stage. Intelligent mechanisms which automatically preserves privacy, without making overburdening  users to define their privacy rules, are yet to be developed. New emerging concepts such as ``collective privacy'' are interesting \cite{Squicciarini2009Collective}.

{\bf Harnessing Quantum.} Exploiting recent advances in quantum computing could enable this giant leap where certain calculations can be done much faster than any classical computer could ever hope to do. For \ac{VR}, quantumness could be leveraged for: 1) bridging virtual and physical worlds, where the classical notion of locality no longer matters, 2) in terms of computation power, where instead of serial or even parallel computation/processing, quantum allows to calculate/compute high-dimensional objects  in lower-dimensions, exploiting entanglement and superposition. This can be instrumental for self-driving vehicles where latency is crucial, therein quantum computing empowers vehicles to recognize and categorize a large number of objects in a real-time manner by solving highly  complex pattern recognition problems on a much faster timescale. 
%
\section{Numerical Results}
\label{sec:numresults}
In this section, in the light of aforementioned challenges, we examine a number of case studies focusing on some of the fundamentals of \ac{AR}/\ac{VR}. Let us suppose that arbitrary number of \ac{AR}/\ac{VR} devices are connected to $M$ fog servers (or base stations) via wireless links with total link capacity of $L_{\mathrm{wi}}$ Mb/s. These fog servers are connected to a cloud computing service (and internet) via backhaul links with total link capacity of $L_{\mathrm{ba}}$ MBit/s. Each \ac{AR}/\ac{VR} device and fog base station have computing capabilities of  $C_{\mathrm{vr}}$ and $C_{\mathrm{fg}}$ GHz respectively, and the cloud has computational power of $C_{\mathrm{cl}}$ GHz. In the numerical setups of the following case studies, the arrival process of \ac{AR}/VR devices  shall follow a Shot Noise Model \cite{Paschos2016Wireless} with a total time period of $T_{\mathrm{max}}$ hours. This model conveniently aims to capture spatio-temporal correlations, where each shot is considered as a \ac{VR} device that stays in the network for a duration of  $T$ ms, and each device has $\mu$ mean number of task requests drawn from a power-law distribution \cite{Leconte2016Placing} with exponent $\alpha$. Requested tasks are computed at different locations of the network, which could be locally at the \ac{VR} device or (edge) fog server or globally at the cloud. Depending on where the requested task is computed, computational and delivery/communication costs are incurred, following power-law distributions parameterized by means $\mu_{\mathrm{co}}$ giga cycles, $\mu_{\mathrm{de}}$ MBit and power-law exponents (or steepness factors) $\alpha_{\mathrm{co}}$ and  $\alpha_{\mathrm{de}}$, respectively. Moreover, computation and delivery of a task incur delays. As a main performance metric, the \emph{immersive experience} is defined as the percentage of tasks which are computed and delivered under a specific deadline, where each deadline is drawn from a power-law distribution with  mean $\mu_{\mathrm{dl}} = 10$ ms and a steepness factor $\alpha_{\mathrm{dl}}$. Such a definition of immersive exeperience is analogous to coverage/outage probability used in the literature, where the aforementioned target task deadline with mean of $10$ ms is imposed for users/humans to avoid noticing lag (no more than $13$ ms in reality \cite{Potter2014Detecting}). A set of default parameters\footnote{$M = 4$,
$L_{\mathrm{ba}} = 512$ Mb/s,
$L_{\mathrm{wi}} = 1024$ Mb/s,
$C_{\mathrm{vr}} = 4 \times 3.4$ GHz,
$C_{\mathrm{fg}} = 128 \times 4 \times 3.4$ GHz, 
$C_{\mathrm{cl}} = 1024 \times 4 \times 3.4$ GHz
$T_{\mathrm{max}} = 1$ hour,
$T = 4$ ms,
$\mu = 4$ tasks,
$\alpha = 0.8$,
$\mu_{\mathrm{co}} = 100$ Giga cycles,
$\alpha_{\mathrm{co}} = 0.48$,
$\mu_{\mathrm{de}} = 100$ MBit,
$\alpha_{\mathrm{de}} = 0.48$,
$\mu_{\mathrm{dl}} = 10$ ms,
$\alpha_{\mathrm{dl}} = 0.48$.
} is considered throughout the case studies, unless otherwise stated. These parameters are to set such values such that a realistic network with  limited storage, computation, and communication capacities is mimicked.

\subsection{Case study I: Joint resource allocation and computing}
The goal is to maximize  a user's immersive experience  by minimizing a suitable cost function. This optimization problem hinges  on many parameters such as the wireless link between the \ac{VR} device and the server (or cluster of servers), whether the \ac{VR} device is a  human or a robot, network congestion, in-\ac{VR} processing power/storage/memory,  and cost (how much intelligence can be embedded in the \ac{VR} device). 

\begin{figure}[!ht]
\centering
\hspace{-1.0cm}
\begin{tikzpicture}[scale=0.9]
	\begin{axis}[
 		grid = major,
 		legend cell align=left,
 		mark repeat={2},
	    legend columns=2,
		legend entries={
			{VR I ~~~({\bf 37\%}, 33\%, 30\%)},
			{VR II ~~~({\bf 59\%}, 25\%, 16\%)},
			{Fog I \hspace{0.3cm}(30\%, {\bf 37\%}, 33\%)},
			{Fog II \hspace{0.3cm}(16\%, {\bf 59\%}, 25\%)},
			{Cloud I (30\%, 33\%, {\bf 37\%})},
			{Cloud II (16\%, 25\%, {\bf 59\%})}
		},
		legend cell align=left,
	    legend to name=namedmethods4vr1,
	    legend style={font=\small}, 		
 		xlabel={\small Arrival density [shot/ms]},
 		ylabel={\small  Immersive Experience [\%]}]
	
		\addplot [color=cyan!70!black, mark=o, thick]  
		         table [col sep=comma, y expr = \thisrowno{1}*100] {\string"fig1curve1a.csv"};
	 	\addplot [color=cyan!70!black, mark=otimes, thick] 
	 			 table [col sep=comma, y expr = \thisrowno{1}*100] {\string"fig1curve1b.csv"};

		\addplot [color=yellow!70!black,  mark=+, thick]  
				 table [col sep=comma, y expr = \thisrowno{1}*100] {\string"fig1curve2a.csv"};
		\addplot [color=yellow!70!black, mark=star, thick]    
	 			 table [col sep=comma, y expr = \thisrowno{1}*100] {\string"fig1curve2b.csv"}; 
	 			 				 
 		\addplot [color=red!70!black,  mark=square, thick]  
 				 table [col sep=comma, y expr = \thisrowno{1}*100] {\string"fig1curve3a.csv"};
 		\addplot [color=red!70!black, mark=triangle, thick]
 				 table [col sep=comma, y expr = \thisrowno{1}*100] {\string"fig1curve3b.csv"}; 	      	
 	\end{axis}
\end{tikzpicture}
\\
\hspace{0.65cm}
\ref{namedmethods4vr1}
\caption{\small Evolution of the immersive experience with respect to the load, with different configurations of VR, Fog, and Cloud-centric computations: 
\underline{VR I} ($C_{\mathrm{vr}} = 2 \times 3.2$ GHz),
\underline{VR II} ($C_{\mathrm{vr}} = 1 \times 3.2$ GHz), 
\underline{Fog I} ($C_{\mathrm{fg}} = 256 \times 4 \times 3.4$ GHz, $L_{\mathrm{wi}} = 1024$ Mb/s),
\underline{Fog II} ($C_{\mathrm{fg}} = 16 \times 4 \times 3.4$ GHz, $L_{\mathrm{wi}} = 256$ Mb/s),
\underline{Cloud I} ($C_{\mathrm{cl}} = 1024 \times 4 \times 3.4$ GHz, $L_{\mathrm{ba}} = 512$ Mb/s),
\underline{Cloud II} ($C_{\mathrm{cl}} = 128 \times 4 \times 3.4$ GHz, $L_{\mathrm{ba}} = 16$ Mb/s). The triple $(., ., .)$ given in the legend represents the percentage of tasks computed at the \ac{VR} devices, fog base stations and cloud, respectively. \vspace{-0.4cm}}
\label{fig:cs1jointresource}
\end{figure}
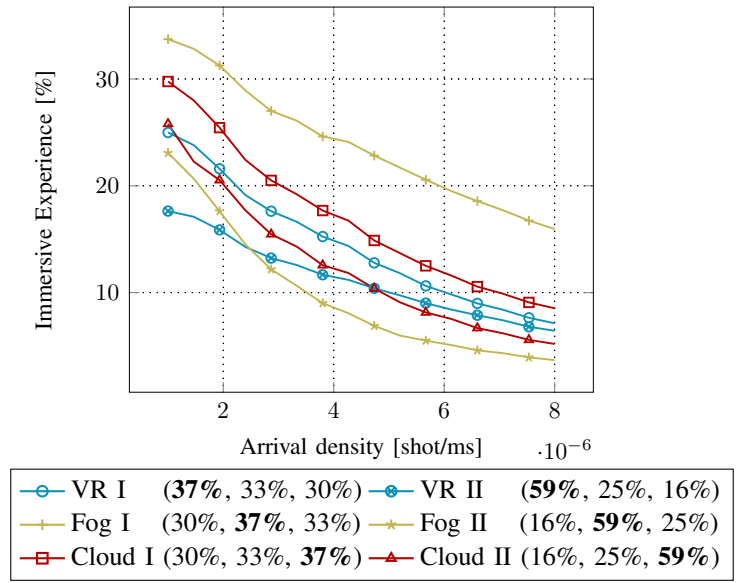

The evolution of the immersive experience with respect to the arrival density of \ac{VR} devices is depicted in Fig. \ref{fig:cs1jointresource}. The tasks are computed at three different places (i.e., locally at \ac{VR} devices, fog base stations or globally at the cloud) with different percentages, in order to show the possible gains. As the arrival density of tasks increases, one can easily see that the immersive experience decreases due to the limited computing and communication resources in the network causing higher delays. In this configuration with $10$ ms average delay deadline/requirement, computing at the fog base stations  outperforms other approaches as seen in the figure. For instance, with an arrival density of $~0.42$ \ac{VR} devices per msec, Fog I provides  $16\%$ more immersive experience gains as compared to other configurations. However, there exist regimes where \ac{VR}-centric computations outperform others (i.e., VR II vs Fog II), especially for higher task arrival densities. The results indicate the need for a principled framework that jointly allocate resources (radio, computing) in various network locations subject to latency and reliability constraints.

\subsection{Case study II: Proactive vs. reactive computing}
Related to the local vs. edge computing challenge in Section \ref{sec:research:computing}, the goal of cloud service providers is to enable tenants to elastically scale resources to meet their demands. While running cloud applications, a tenant aiming to minimize her/his cost function is often challenged with crucial tradeoffs. For instance, upon each arrival of a task, an application can either choose to pay for \ac{CPU} to compute the response, or pay for cache storage to store the response to reduce future compute costs. Indeed, a reactive computing approach would wait until the task request reaches the server for computation, whereas the proactive computing approach  proactively leverages that fact that several requests/queries will be made for the same computation, and thus it stores the result of the computation in its cache to avoid recomputing the query at each time instant. This fundamental observation is analysed next.

\begin{figure}[!ht]
\centering
\hspace{-1.0cm}
\begin{tikzpicture}[scale=0.9]
	\begin{axis}[
 		grid = major,
 		legend cell align=left,
 		mark repeat={2},
	    legend columns=3,
		legend entries={
			Reactive ~L,
			Reactive ~M,
			Reactive ~H,
			Proactive L,
			Proactive M,
			Proactive H		
		},
		legend cell align=left,
	    legend to name=namedmethods4vr2,
	    legend style={font=\small}, 		
 		xlabel={\small Proactivity [\%]},
 		ylabel={\small  Immersive Experience [\%]}]
	
		\addplot [color=red!70!black, mark=o, thick]  
		         table [col sep=comma, y expr = \thisrowno{1}*100, x expr = \thisrowno{0}*100]{\string"fig2curve1a.csv"};
	 	\addplot [color=red!70!black, mark=+, thick] 
	 			 table [col sep=comma, y expr = \thisrowno{1}*100, x expr = \thisrowno{0}*100]{\string"fig2curve1b.csv"};   	
	 	\addplot [color=red!70!black, mark=square, thick] 
	 			 table [col sep=comma, y expr = \thisrowno{1}*100, x expr = \thisrowno{0}*100]{\string"fig2curve1c.csv"};  	 			 
 		 		
 		\addplot [color=cyan!70!black, mark=otimes, thick]  
 				 table [col sep=comma, y expr = \thisrowno{1}*100, x expr = \thisrowno{0}*100]{\string"fig2curve2a.csv"};
 		\addplot [color=cyan!70!black, mark=star, thick]  
 				 table [col sep=comma, y expr = \thisrowno{1}*100, x expr = \thisrowno{0}*100]{\string"fig2curve2b.csv"}; 				 
 		\addplot [color=cyan!70!black, mark=triangle, thick] 
 				 table [col sep=comma, y expr = \thisrowno{1}*100, x expr = \thisrowno{0}*100]{\string"fig2curve2c.csv"}; 	      	
 	\end{axis}
\end{tikzpicture}
\vspace{0.2cm}
\scriptsize
\\
\hspace{-0.0cm}
\ref{namedmethods4vr2}
\caption{\small Evolution of the immersive experience with respect to the proactivity. Low (L), medium (M), and high (H) homogeneity settings for reactive and proactive computation of tasks at the fog servers are considered:
\underline{Reactive L} ($\alpha = 0.1$, $L_{\mathrm{ba}} = 64$ Mb/s)
\underline{Reactive M} ($\alpha = 0.6$, $L_{\mathrm{ba}} = 64$ Mb/s)
\underline{Reactive H} ($\alpha = 0.8$, $L_{\mathrm{ba}} = 64$ Mb/s)
\underline{Proactive L} ($\alpha = 0.1$, $L_{\mathrm{ba}} = 64$ Mb/s)
\underline{Proactive M} ($\alpha = 0.6$, $L_{\mathrm{ba}} = 64$ Mb/s)
\underline{Proactive H} ($\alpha = 0.8$, $L_{\mathrm{ba}} = 64$ Mb/s). The place of computations for all settings is fixed to  $(16\%, 25\%, {\bf 59\%})$.\vspace{-0.4cm}}
\label{fig:cs2proactive}
\end{figure}
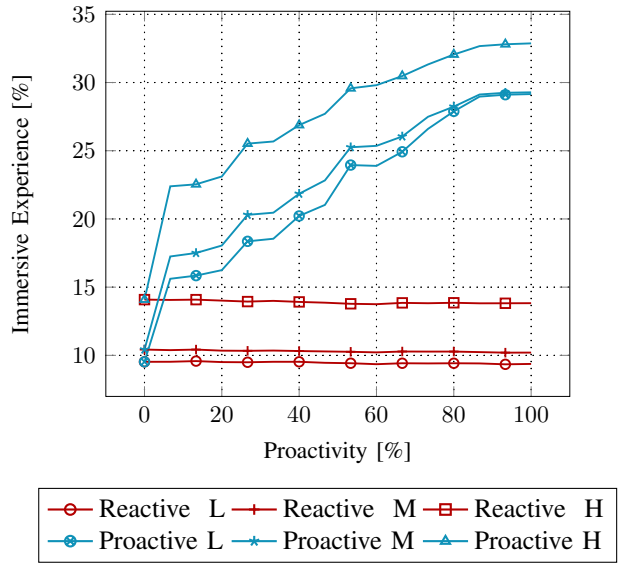

The evolution of the immersive experience with respect to the level of proactivity at the fog base stations is shown in Fig. \ref{fig:cs2proactive}. We assume proactive settings with storage size of $S = 0\%$ in case of zero proactivity and $S = 100\%$ in case of full proactivity, whereas the computation results of popular tasks are cached in the fog base stations for a given storage size. As seen in the figure, proactivity substantially increases the immersive experience, and further gains are obtained when the computed tasks are highly homogeneous (i.e., Proactive H). The gains in the reactive approaches remain  constant as there is no proactivity, whereas a slight improvement in highly homogeneous case (i.e., Reactive H) is observed due to the homogeneous tasks that are prone to less fluctuations in deadlines. As an example, $80\%$ of proactivity in Proactive H yields higher gains up to $22\%$ as compared to Reactive L. This underscores the compelling need for proactivity in VR systems.

\subsection{Case study III: AR-enabled Self-Driving Vehicles}
Self-driving or autonomous vehicles represent one of the most important use case for 5G where latency, bandwidth and reliability are prime concerns. Self-driving vehicles  need to exchange  information  derived  from  multi-resolution maps created using their local sensing modalities (radar, lidar, or cameras), extending their  visibility  beyond  the  area directly  sensed  by  its  own  sensors.  The problems facing the vehicles are many-fold: 1) how to control the size of the message (payload) exchanged with other  vehicles based on traffic load, interference, and other contextual information; 2)  how to control the content of the message (at what granularity should a given object be included in the message, the most popular object? the least requested object? at what timeliness, etc.); 3) how to recognize objects and patterns reliably and in real-time?

\begin{figure}[!ht]
\centering
\hspace{-1.0cm}
\begin{tikzpicture}[scale=0.9]
	\begin{axis}[
 		grid = major,
 		legend cell align=left,
 		mark repeat={2},
	    legend columns=2,
		legend entries={
			{VR R \hspace{0.31cm}({\bf 50\%}, 30\%, 20\%)},
			{VR P \hspace{0.31cm}({\bf 50\%}, 30\%, 20\%)},
			{Fog R \hspace{0.26cm}(20\%, {\bf 50\%}, 30\%)},
			{Fog P \hspace{0.26cm}(20\%, {\bf 50\%}, 30\%)},
			{Cloud R (20\%, 30\%, {\bf 50\%})},
			{Cloud P (20\%, 30\%, {\bf 50\%})}
		},
		legend cell align=left,
	    legend to name=namedmethods4vr3,
	    legend style={font=\small}, 		
 		xlabel={\small Channel Congestion [\%]},
 		ylabel={\small  Immersive Experience  [\%]}]
	
 		\addplot [color=cyan!70!black, mark=o, thick]  
 				 table [col sep=comma, y expr = \thisrowno{1}*100, x expr = \thisrowno{0}*100] {\string"fig3curve1a.csv"};		
 		\addplot [color=cyan!70!black, mark=otimes, thick]
 				 table [col sep=comma, y expr = \thisrowno{1}*100, x expr = \thisrowno{0}*100] {\string"fig3curve1b.csv"};

 		\addplot [color=yellow!70!black, mark=+, thick]  
 				 table [col sep=comma, y expr = \thisrowno{1}*100, x expr = \thisrowno{0}*100] {\string"fig3curve2a.csv"};			  	
 		\addplot [color=yellow!70!black, mark=star, thick]
 				 table [col sep=comma, y expr = \thisrowno{1}*100, x expr = \thisrowno{0}*100] {\string"fig3curve2b.csv"};

 		\addplot [color=red!70!black, mark=square, thick]  
 				 table [col sep=comma, y expr = \thisrowno{1}*100, x expr = \thisrowno{0}*100] {\string"fig3curve3a.csv"};	     				 
 		\addplot [color=red!70!black, mark=triangle, thick]
 				 table [col sep=comma, y expr = \thisrowno{1}*100, x expr = \thisrowno{0}*100] {\string"fig3curve3b.csv"}; 	      	
 	\end{axis}
\end{tikzpicture}
\vspace{0.2cm}
\scriptsize
\\
\hspace{0.65cm}
\ref{namedmethods4vr3}
\caption{\small Evolution of the immersive experience with respect to the channel congestion, where fully reactive (R) and proactive (P) configurations of VR, Fog, Cloud-centric computation are considered. Fully reactive configuration has $S = 0$\%, $L_{\mathrm{ba}} = 64$ Mb/s; and the proactive configuration has $S = 80$\%, $L_{\mathrm{ba}} = 64$ Mb/s.
\vspace{-0.4cm}}
\label{fig:cs3selfdriving}
\end{figure}
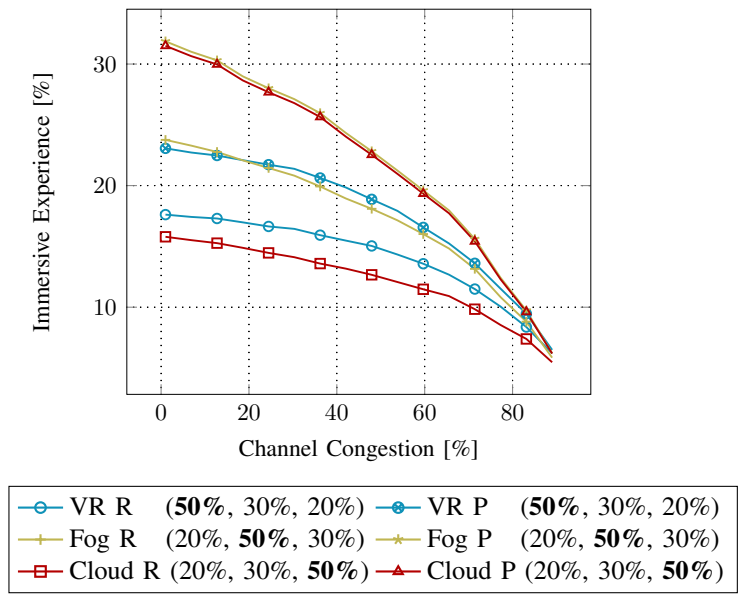

The evolution of the immersive experience with respect to the wireless channel link congestion between base stations and AR-enabled self-driving vehicles is depicted in Fig. \ref{fig:cs3selfdriving}. The fact that higher channel congestion degrades the immersive experience in all settings is evident, however, proactivity can still provide additional improvements as compared to the reactive settings. Proactive cloud and fog oriented computation yield gains up to $11\%$ when the congestion is $42\%$. This shows the need of proactivity in self-driving vehicles as well as dynamic placement of computation depending on the \ac{AR}/\ac{VR} network conditions.

Delving into these case studies which show the potential of interconnected \ac{VR}, we finally come to the following question.

\section{Are We Going to Live in the "Matrix"?}\label{sec:idealvr}
One speculative question which can be raised is whether an interconnected \ac{VR} can reach to a maturity level so that  no distinction between real and virtual worlds are made in human perception, making people end up with the following question: \emph{Are we living in a computer simulation?} 

Despite historical debates, several science-fiction movies have been raising similar points (i.e. The Matrix), many philosophical discussions have been carried out \cite{Bostrom2003We}, concepts like "simulated reality"  have been highlighted \cite{Barrow2007Living}, and despite all of these, many technical and scientific challenges remain unclear/unsolved. In the context of \ac{VR}, we call this unreachable phenomenon as \emph{ideal (fully-interconnected) \ac{VR}}. In fact, in the realm of ideal \ac{VR}, one might think of living in a huge computer simulation with zero distinction/switching between real and virtual worlds. In this ideal \ac{VR} environment, the concepts of skin/edge/fog/cloud computing might be merged with concepts like quantum computing.

Indeed, in ideal \ac{VR} with no distinction between real and virtual worlds, we are not aiming to introduce a paradoxical concept and provide recursive arguments with mixture/twist of ideas. Instead, we argue whether we can reach such a user experience with \ac{VR}, therefore achieving an ideal (fully-interconnected) case. Despite the fact that we do not know the exact answer, we keep the ideal \ac{VR} as a reference to all interconnected \ac{VR} systems. Undoubtedly, the future lies in  interconnected \ac{VR}, despite its research and scientific challenges which will continue to grow in importance over the next couple of years.
\bibliographystyle{IEEEtran}
\bibliography{references}
\end{document}